\documentclass[eqsecnum,nofootinbib,12pt]{revtex4}
   \usepackage{bm}
\usepackage{amsthm,amscd}
\usepackage{mathrsfs}
\usepackage{verbatim}
\usepackage{amsfonts,amsmath,latexsym,amssymb,bm}
\usepackage[american]{babel}
\usepackage{dsfont}
\usepackage{epsfig}
\usepackage{color,soul}

\newcommand\bea{\begin{eqnarray}}
\newcommand\eea{\end{eqnarray}}
\renewcommand{\mathcal}{\mathscr}
\newcommand{\beq}{\begin{eqnarray}}
\newcommand{\eeq}{\end{eqnarray}}
\newcommand{\nn}{\nonumber}
\newcommand{\com}{\mbox{${\rm I\!\!\!C }$}}
\newcommand{\reals}{\mbox{${\rm I\!R }$}}

\begin{document}
\title{Casimir Effect in the Presence of External Fields}
\author{
Matthew Beauregard\footnote{Electronic address: Matthew\textunderscore Beauregard@Baylor.edu},
Klaus Kirsten\footnote{Electronic address: Klaus\textunderscore Kirsten@Baylor.edu} and Pedro Morales\footnote{Electronic address: Pedro\textunderscore Morales@Baylor.edu}
\thanks{Electronic address: gfucci@nmt.edu}}
\affiliation{Department of Mathematics, Baylor University, Waco, TX 76798 USA
}
\author{Guglielmo Fucci\footnote{Electronic address: fuccig@ecu.edu} }
\affiliation{Department of Mathematics, East Carolina University, Greenville, NC 27858 USA}

\date{\today}
\vspace{2cm}
\begin{abstract}

In this work the Casimir effect is studied for scalar fields in the presence of boundaries and under the influence
of arbitrary smooth potentials of compact support. In this setting, piston configurations are analyzed in which
the piston is modeled by a potential. For these configurations, analytic results for the Casimir energy and force are obtained
by employing the zeta function regularization method. Also, explicit numerical results for the Casimir force are provided
for pistons modeled by a class of compactly supported potentials that are realizable as delta-sequences. These results are then generalized to higher dimensional pistons by
considering additional Kaluza-Klein dimensions.

\end{abstract}
\maketitle

\section{Introduction}
In 1948 it was shown by Casimir \cite{casi48-51-793} that the presence of a boundary significantly modifies the vacuum structure of a quantum field. In particular, he computed the pressure between two perfectly conducting parallel plates due to the ground state of an electromagnetic field and determined that an attractive force exists between them. As a result, an enormous amount of literature has been produced which has been focused on the analysis of the effects that the geometry of the boundary and the boundary conditions have on the vacuum structure of quantum fields (see e.g. \cite{milt01b,bord01-353-1,bord09b,eliz95b,kirs02b,milo94b,most97b} and references therein). In the vast majority of cases, however, research has been centered on the analysis of quantum systems endowed with ideal boundary conditions. The reason for such polarized interest lies in the fact that the eigenvalues and the eigenfunctions of the relevant Laplace type operator with ideal boundary conditions  are often explicitly known. This makes the analysis of quantum vacuum effects, as a result, less involved.

Ideal boundary conditions, although of crucial theoretical importance for understanding important aspects of quantum vacuum effects, do not always provide an accurate description of physical systems. For instance, it is well known that a neutral metal plate does not constitute an impenetrable boundary for all the frequencies associated with a quantum field. For this reason different methods have been suggested in order to construct models that more closely characterize physical properties of real materials.

One of the approaches used to more precisely describe quantum systems relies on the replacement of ideal boundary conditions with a boundary modeled by a potential. The rationale behind this idea is that it is possible to describe physical characteristics of a boundary by choosing an appropriate potential function. This process reduces to substituting a geometric boundary of the system with a non-dynamical external field. Along these ideas, the analysis of the Casimir effect in the setting of potentials modeling a boundary have appeared, for example, in \cite{acto95-52-3581,eliz97-30-5393}.

It is worth mentioning that the importance of external potentials in quantum systems is not confined to the description of a non-ideal boundary. In fact, they play an important role in describing background potentials resulting from classical solutions like monopoles \cite{hoof74-79-276,poly74-20-194}, sphalerons \cite{klin84-30-2212} and electroweak Skyrmions \cite{gips81-183-524,gips84-231-365,ambj85-256-434,eila86-56-1331,frie77-15-1694,frie77-16-1096,skyr61-260-127,skyr62-31-556,adki83-228-552}.

Furthermore, the formalism needed to study the vacuum energy of scalar fields under the influence of integrable potentials in unbounded Euclidean space has been developed in \cite{bord96-53-5753,dunn06-39-11915,dunn09-42-075402}. It is the purpose of the current work to extend this analysis to include finite spatial volumes and additional Kaluza-Klein dimensions.

The models considered in this paper can be formally reduced to the analysis of the Casimir effect on a one-dimensional finite interval with an arbitrary, sufficiently smooth, potential. Assuming two additional free dimensions this represents a piston configuration in flat space, where the piston is modeled by a potential with a sharply concentrated support. The arbitrary additional dimensions are represented by a smooth Riemannian manifold ${\cal N}$ with or without boundary, which describes the cross-section of the piston. Piston configurations have received significant interest in recent years as they are often free of divergences which allows for an unambiguous prediction of forces \cite{cava04-69-065015}. Prior research has concentrated on infinitely thin pistons where different types of ideal boundary conditions were imposed; see, e.g., \cite{eliz09-79-065023,fucci12a,hert05-95-250402,kirs09-79-065019,mara07-75-085019}. As previously mentioned, this work differs substantially from previous ones in that infinitely thin pistons are replaced by potentials modeling pistons of finite thickness.

In this paper the zeta function regularization method is used, in which the Casimir energy and force are given by contour integrals that involve boundary values of unique solutions to initial value problems. This particular approach has been successfully applied to the evaluation of functional determinants \cite{kirs03-308-502,kirs04-37-4649}, for which explicit results can be obtained through purely analytical means and can be represented in closed form. In order to compute the Casimir energy and the corresponding force, however, numerical integration through suitable quadrature methods is necessary.

The outline of the paper is as follows. In Section \ref{onedim} the spectral zeta function associated with a boundary value problem in one dimension is represented in terms of a contour integral and the formalism needed to perform the analytic continuation to a suitable domain in the complex plane is developed. The obtained meromorphic extension is exploited to analyze the Casimir energy for a massive scalar field on an interval under the influence of a background potential. The Casimir force on the piston is found for cases where the potential mimics a piston; see Equation (\ref{2.16}). In Section \ref{Sec:Examples} numerical results for the Casimir force are given for various potentials constructed from smooth, compactly supported functions, for which each can be realized as a delta-sequence. In Section \ref{hdp}, additional Kaluza-Klein dimensions are included into the analysis and the Casimir energy is determined. In this setting, results for the Casimir force are provided for cases in which the additional dimensions are chosen to be either $\mathbb{R}$ or $\mathbb{R}^{2}$. In the Conclusions, the
results are summarized and additional studies along the lines of this work are outlined.

\section{Massive scalar field on a one-dimensional interval}\label{onedim}
In this section a non-selfinteracting massive scalar field is considered on the interval $I = [0,L]\subset\mathbb{R}$ under the influence of a smooth background potential $V(x)$. The one-particle energy eigenvalues $\lambda_\ell$ of this system are determined by the differential equation
\beq
\left( - \frac{d^2}{dx^2} + V(x) + m^2 \right) \phi_\ell (x) = \lambda_\ell \phi_\ell (x)\;, \label{2.1}
\eeq
augmented by boundary conditions which, for definiteness, are chosen to be of Dirichlet type
\beq
\phi_\ell (0)  = \phi_\ell (L) =0\;.\label{2.2}
\eeq
In what follows, the mass parameter $m$ is assumed large enough so that all the eigenvalues of the problem (\ref{2.1}) and (\ref{2.2}) are strictly positive.

The spectral zeta function associated with the above boundary value problem is defined by
\beq
\zeta_I (s) = \sum_{\ell =0} ^\infty \lambda_\ell ^{-s}\;, \label{2.3}
\eeq
and, due to the
asymptotic behavior of the eigenvalues, it is convergent in the semi-plane $\Re(s) > 1/2$. In the framework of zeta function regularization, the Casimir energy $E_{Cas}$ of the system is encoded in the value of $\zeta_I (s)$ at $s=-1/2$ \cite{bord09b,eliz95b,kirs02b}. More precisely
\beq
E_{Cas} = \lim_{\alpha \to 0} \frac {\mu^{2\alpha}} 2 \zeta_I \left( \alpha - \frac 1 2 \right)\;, \label{2.4}
\eeq
where $\mu$ represents an arbitrary parameter with the dimension of a mass.
Since the point $s=-1/2$ does not belong to the domain of convergence of (\ref{2.4}),
it is necessary to perform the analytic continuation of the series in Equation (\ref{2.3}) to a neighborhood of $s=-1/2$. Here, the strategy employed consists in rewriting the series (\ref{2.3}) as a contour integral using Cauchy's residue theorem \cite{conw78b} and then utilizing the obtained integral representation as a starting point for the analytic continuation. This technique has been successfully used for the calculation of functional determinants in the setting described by Equation (\ref{2.1}) in \cite{kirs03-308-502,kirs04-37-4649} .

Following the ideas developed in \cite{fucci12,kirs03-308-502,kirs04-37-4649}, instead of the boundary value problem (\ref{2.1}) and (\ref{2.2}), the following equivalent {\it initial value problem} is to be considered,
\beq
\left( - \frac {d^2}{dx^2} + V(x) +m^{2}\right) u_\nu (x) = \nu^2 u_\nu (x)\;, \quad \quad u_\nu (0) =0\;, \quad u_\nu ' (0) =1\;, \label{2.5}
\eeq
where $\nu \in \com.$ The eigenvalues $\lambda_\ell$ of the original eigenvalue problem (\ref{2.1}) with Dirichlet boundary conditions (\ref{2.2}) are then determined as solutions to the transcendental equation
\beq
u_\nu (L) =0\;.\label{2.6}
\eeq
Let us point out that the solutions to Equation  (\ref{2.5}) are uniquely determined and define an analytic function of $\nu$.

For $\Re (s) > 1/2$, the zeta function (\ref{2.3}) can be represented in terms of a contour integral as \cite{fucci12}
\beq
\zeta _I (s) = \frac 1 {2\pi i} \int\limits_\gamma d\nu (\nu^2 + m^2 ) ^{-s} \frac d {d\nu} \ln u_\nu (L)\;, \label{2.7}
\eeq
where the contour $\gamma$ encloses all solutions to Equation (\ref{2.6}), assumed to be on the positive real axis,
in the counterclockwise direction. A deformation of the integration contour to the imaginary axis in (\ref{2.7})
leads to the following expression
\beq
\zeta _I (s) = \frac{\sin (\pi s)} \pi \int\limits_m^\infty dk (k^2-m^2)^{-s} \frac d {dk} \ln u_{ik} (L)\;, \label{2.8}
\eeq
which is now valid for $1/2 < \Re (s) < 1$. The analytic continuation of (\ref{2.8}) to the region $\Re(s)\leq 1/2$ is
obtained by adding and subtracting the large-$k$ asymptotic behavior of the solution $u_{ik} (L)$ \cite{fucci12}.

The needed asymptotic behavior can be determined by applying a standard WKB technique to the unique solution of the
initial value problem
\beq
\left( - \frac{d^2} {dx^2} + V(x) + k^2 \right) u_{ik} (x) =0\;, \quad \quad u_{ik} (0) =0\;, \quad u_{ik} ' (0) =1\;.\label{2.9}
\eeq
Although in the process of analytic continuation one only needs to be concerned with the exponentially growing part for large $k$,
at this stage it is important to take into account both the exponentially growing and the exponentially decaying contributions in order to be able to
correctly impose the initial condition in (\ref{2.9}). Following, e.g., \cite{bend10b,mill06b}, it is convenient to introduce the auxiliary function
\beq
S(x,k) = \partial _x \ln \psi_k (x)\;, \label{w1}
\eeq
where $\psi_k (x)$ satisfies
\beq
\left( - \frac {d^2}{dx^2} + V(x) + k^2 \right) \psi _k (x) =0. \label{2.10}
\eeq
By using (\ref{w1}) in (\ref{2.10}), it is not very difficult to show that $S(x,k)$ satisfies the differential equation
\beq
S ' (x,k) = k^2 + V(x) - S^2 (x,k)\;, \label{2.11}
\eeq
where, here and in the rest of this work, the prime indicates differentiation with respect to the variable $x$.
The asymptotic expansion of $S(x,k)$ for large $k$ can be written in the form
\beq
S(x,k) = \sum _{i=-1}^\infty k^{-i} S_i (x)\;,
\eeq
where
the asymptotic orders $S_i (x)$ are recursively determined  by
\beq
S_{-1}(x) &=& \pm 1\;, \quad S_0 (x) =0\;, \quad S_1 (x) = \pm \frac{V(x)} 2\;, \label{2.12}\\
S_{i+1} (x) &=& \mp \frac 1 2 \left(S_i ' (x) + \sum_{j=0}^i S_j (x) S_{i-j} (x) \right)\;.\nn
\eeq

The two different signs in (\ref{2.12}) correspond to the exponentially growing and decaying solutions $\psi_k (x)$ of (\ref{2.10}). Let $S^{\pm } (x,k)$ denote the solutions of (\ref{2.11}) corresponding to the different signs, the associated solutions of (\ref{2.10}) have the form
\beq
\psi_k ^\pm (x) = A^\pm \exp \left\{ \int\limits_0^x dt \,\, S^\pm (t,k) \right\}\;.
\eeq
The original function of interest, namely $u_{ik} (x)$, is obtained as a linear combination
\beq \label{w3}
u_{ik} (x) = A^+ \exp \left\{ \int\limits_0^x dt \,\, S^+ (t,k) \right\} + A^- \exp \left\{ \int\limits_0^x dt \,\, S^- (t,k) \right\}\;,
\eeq
where the arbitrary coefficients $A^{+}$ and $A^{-}$ can be found by imposing the initial condition in (\ref{2.9}) and they read
\beq \label{w4}
A^+ = - A^-, \quad \quad A^+ = \frac 1 {S^+ (0,k) - S^- (0,k)}\;.
\eeq
By using the result (\ref{w4}) in the expression (\ref{w3}) the large-$k$ behavior of $u_{ik} (L)$ can be found to be
\beq
u_{ik} (L) = \frac 1 {S^+ (0,k) - S^- (0,k)} \exp \left\{ \int\limits_0^L dt \,\, S^+(t,k) \right\} \big(1+ E(k)\big)\;,
\eeq
where $E(k)$ denotes exponentially decreasing terms as $k\to\infty$.
For the relevant quantity in the integral (\ref{2.8}),
\beq \label{w61}
\ln u_{ik} (L) &=& - \ln \left( S^+ (0,k) - S^- (0,k)\right) + \int\limits_0^L dt \,\, S^+(t,k)  \nn\\
&=& - \ln (2k) + k L+ \sum_{j=0}^\infty d_j k^{-j}\;,
\eeq
where exponentially small terms have been omitted and the coefficients $d_j$ are defined as
\beq\label{w5}
d_{2j+1}=\int\limits_{0}^{L}dx\,S_{2j+1}^{+}(x)\;,
\eeq
and
\beq
d_{2j}=\int\limits_{0}^{L}dx\,S_{2j}^{+}(x)-\Omega_{j}(0)\;,
\eeq
with $\Omega_{j}(0)$ defined through the cumulant expansion
\beq
\ln\left[1+\sum_{k=1}^{\infty}\frac{S^{+}_{2k-1}(0)}{z^{2k}}\right]\simeq\sum_{i=1}^{\infty}\frac{\Omega_{i}(0)}{z^{2i}}\;.
\eeq
For completeness, the first six coefficients $d_j$ are explicitly given by
\beq
d_0&=&0\;,\quad\quad d_1=\frac{1}{2}\int\limits_0^L dt\,\,V(t)\;,\quad\quad d_2 = - \frac 1 4 [V(L) + V(0) ]\;,\nn\\
d_3&=&\frac{1}{8}[V'(L)-V'(0)]-
\frac{1}{8}\int\limits_0^Ldt\,\, V^2(t)\;,\nn\\
d_4&=&-\frac{1}{16}[V''(L)+V''(0)]+\frac{1}{8}[V^2(L)-V^2(0)]\;,\label{dsubi}\\
d_5&=&\frac{1}{32}[V^{(3)}(L)-V^{(3)}(0)]-\frac{5}{32}[V(L)V'(L)-V(0)V'(0)]+\frac{1}{16}
\int\limits_0^Ldt\,\,V^3(t)\nn\\
&~&-\frac{1}{32}\int\limits_0^Ldt\,\,V(t)V''(t)\;.\nn
\eeq
It is clear that by using (\ref{2.12}) and the definition (\ref{w5}), an arbitrary number of coefficients can be determined by using an algebraic computer program.

By adding and subtracting from the integral representation (\ref{2.8}) the leading $N+1$ terms in the asymptotic expansion (\ref{w6}), the zeta function can be represented as a sum of two terms
\beq\label{w6}
\zeta_I (s) = \zeta _I ^{(f)} (s) + \zeta _I ^{(as)} (s)\;,
\eeq
where \cite{fucci12}
\beq
\zeta_I ^{(f)} (s) = \frac{\sin \pi s} \pi \int\limits_{m} ^\infty dk \,\, (k^2 - m^2)^{-s} \frac d {dk} \left\{ \ln u_{ik} (L) -kL + \ln (2k) -\sum_{j=0}^N d_j k^{-j} \right\}\;, \label{2.14}
\eeq
and
\beq
\zeta_I ^{(as)} (s) =  \frac{\sin \pi s} \pi \int\limits_{m} ^\infty dk \,\, (k^2 - m^2)^{-s} \frac d {dk} \left\{ kL - \ln (2k) +\sum_{j=0}^N d_j k^{-j} \right\}\;. \label{2.15}
\eeq
By construction, the function $\zeta_I^{(f)} (s)$ is analytic in the region $\Re (s) >-(N+1)/2$ and the meromorphic structure of $\zeta_I ^{(as)} (s)$ is made manifest once the $k$-integration is performed. More explicitly,
\beq \label{w7}
\zeta_I ^{(as)} (s) = \frac 1 {2 \Gamma (s)} \left\{ \frac{L \Gamma \left( s- \frac 1 2 \right)}{\sqrt \pi} m^{1-2s}
- \Gamma (s) m ^{-2s} - \sum_{j=1}^N j d_j \frac{ \Gamma \left( s+ \frac j 2 \right)}{\Gamma \left( 1 + \frac j 2 \right)} m^{-j-2s} \right\}\;,
\eeq
where it is clear now that $\zeta_I ^{(as)} (s)$ represents a meromorphic function in the entire complex plane possessing only simple poles.
The expression (\ref{w6}) together with (\ref{2.14}) and (\ref{2.15}) represents the desired analytic continuation of the spectral zeta function (\ref{2.7}).

For the purpose of computing the Casimir energy it is sufficient to choose $N=1$ in the above expressions for $\zeta_I ^{(f)} (s)$ and  $\zeta_I ^{(as)} (s)$.
In this case, $\zeta_I ^{(f)} (s)$ is analytic for $\Re(s)>-1$ and one can simply set $s=-1/2$ in (\ref{2.14}) to obtain
\beq\label{w8}
\zeta ^{(f)} _I(-1/2) &=& - \frac 1 \pi \int\limits_{m } ^\infty dk \,\, (k^2 - m^2)^{1/2} \frac d {dk} \left\{ \ln u_{ik} (L) - kL + \ln (2k) -  \frac{d_1}{ k} \right\} ,
\eeq
while, in the neighborhood of $s=-1/2$, (\ref{w7}) gives
\begin{eqnarray}\label{w9}
\lefteqn{\zeta_I^{(as)} (-1/2+\alpha ) = \frac 1 \alpha \frac{ 2 d_1 + Lm^2}{4\pi}}\nn\\
&+& \frac 1 {4\pi} \left[ - 2m\pi + Lm^2 (\ln 4-1) + 4d_1 (\ln 2-1) - 2 (2d_1 + Lm^2) \ln m\right] + O(\alpha )\;.
\end{eqnarray}
The explicit form of the Casimir energy for this system easily follows by substituting (\ref{w8}) and (\ref{w9})
in the following expression
\begin{equation}\label{w10}
  E_{Cas}=\frac{1}{2}\textrm{FP}\zeta_{I}\left(-\frac{1}{2}\right)+\frac{1}{2}\left(\frac{1}{\alpha}+\ln\mu^{2}\right)\textrm{Res}\,\zeta_{I}\left(-\frac{1}{2}\right)+O(\alpha)\;,
\end{equation}
obtained by expanding (\ref{2.4}) about $\alpha= 0$. In the above formula and throughout the rest of this paper $\textrm{Res}$ denotes the residue of the function and $\textrm{FP}$ its finite part.
It is evident from (\ref{w9}) and (\ref{w10}) that the Casimir energy is, in general, not well defined because of the presence of the term $\textrm{Res}\,\zeta_{I}(-1/2)$.
In order to overcome this problem, the system is interpreted in terms of a piston configuration where the piston itself is modeled by the potential $V(x)$.
For this purpose, the potential $V(x)$ is assumed to have compact support within the interval $[0,L]$. More precisely, it is assumed that $V(x)$ does not vanish for $x\in [a-\epsilon , a+\epsilon] \subset [0,L]$. According to this description, the point $x=a$ represents the position of the piston.

The asymptotic terms $d_{j}$ are expressed either in terms of boundary values of $V(x)$ and its derivatives or as an integral of $V(x)$ and its derivatives, therefore they are independent of the position $a$. It follows from the previous remarks that the Casimir force on the piston, defined in terms of the Casimir energy as
\beq
F_{Cas}=-\frac{\partial}{\partial a}E_{Cas}\;,
\eeq
is a well defined quantity since $\textrm{Res}\,\zeta_{I}(-1/2)$ is, in this setting, independent of $a$. Hence, the explicit expression for the force is
\beq
F_{Cas}(a) = \frac 1 {2\pi }  \int\limits_{m} ^\infty dk \,\,
(k^2 - m^2)^{1/2} \,\, \frac \partial {\partial a} \frac \partial {\partial k} \ln u_{ik} (L)\;.\label{2.16}
\eeq
It is worth pointing out that according to the above formula
the magnitude and direction of the force is encoded in the boundary values of
the solution to an initial value problem associated with an ordinary differential equation and no additional information is necessary.

Despite the simplicity of (\ref{2.16}) information about the behavior of $F_{Cas}$ as a function of $a$ can only be obtained through numerical integration techniques since the solution of (\ref{2.9}) is not explicitly known for an arbitrary $V(x)$. In the following section the analysis of $F_{Cas}$ is provided for different types of potentials constructed from smooth, compactly supported functions.

\section{Examples: Gaussian Potentials}\label{Sec:Examples}

It is clear, from (\ref{2.16}), that in order to extract information about the Casimir force $F_{Cas}$ the evaluation of $u_{ik} (L)$ is necessary.
An immediate numerical concern is that solutions to the differential expression
(\ref{2.9}) for large $k$ contain an exponentially increasing term of the type $e^{kx}.$ For this reason, to ensure accuracy in the numerical evaluation, discretization sizes must be chosen to be sufficiently small.  This restriction is computationally costly but can be circumvented with relative ease in the following manner.  Let $u_{ik} (x) = e^{kx} \varphi_{ik} (x)$ in (\ref{2.9}).  The newly introduced function $\varphi_{ik} (x)$ satisfies the initial value problem
 \beq
 \left( - \frac {d^2} {dx^2} - 2k \frac d {dx} + V(x) \right) \varphi_{ik} (x) =0\;, \quad \quad \varphi_{ik } (0) =0\;, \quad \varphi _{ik} ' (0) =1\;.\label{3.1}
 \eeq
The Casimir force in (\ref{2.16}) can, therefore, be written as
\begin{equation}
  F_{Cas}(a)=\frac 1 {2\pi }  \int\limits_{m} ^\infty dk \,\,
(k^2 - m^2)^{1/2} \,\, \frac \partial {\partial a} \frac \partial {\partial k}\left[e^{kx} \ln \varphi_{ik} (L)\right]\;,
\end{equation}
and since the exponentially growing term does not depend on $a$, this simplifies to
\beq
F_{Cas}(a) =  \frac 1 {2\pi }  \int\limits_{m} ^\infty dk \,\,
(k^2 - m^2)^{1/2} \,\,\frac \partial {\partial a} \frac \partial {\partial k} \ln \varphi_{ik} (L)\;.\label{3.2}
\eeq
The expression (\ref{3.2}) is now suitable for a numerical evaluation since the exponentially growing term has been dealt with analytically
and, as a result, stringent tolerances on the discretization sizes have been alleviated.

In the next subsections, the results for the Casimir force on Gaussian potentials centered at $a$ and with support of extension $\epsilon$ are presented. More specifically,  background potentials of the form are considered,
\beq \label{w11}
V(x) = \left\{ \begin{array} {ll}\eta \frac{\exp \left( \frac{-(x-a)^2}{\epsilon ^2 - (x-a)^2} \right)}{\left|\int\limits_0^1 \eta \exp \left( \frac{-(y-a)^2}{\epsilon ^2 - (y-a)^2} \right) dy \right|} &
\mbox{for }|x-a| < \epsilon, \\
0 & \mbox{otherwise}\;. \end{array} \right.
\eeq
For simplicity, in the following examples $m=0$, $L=1$, and $\epsilon = 10^{-4}.$

In order to obtain the Casimir force on the piston as a function of the position $a$ from the expression (\ref{3.2}) an adaptive second order Runge-Kutta method is used to integrate the differential equation (\ref{3.1}), with a potential of the form (\ref{w11}).  This approximates the value of $\varphi_{ik}(x)$ at $x=L$.  For potentials of the type (\ref{w11}) this method can be shown to be convergent. More importantly, the discretization size can be determined such that the error in the approximation is within a user-specified tolerance. Upon successful calculation of $\varphi_{ik}(L)$, standard second order centered differences are used to approximate the necessary derivatives in the integral formulation for the Casimir force.

In the next step, the integral over the finite interval $(m,M)$ is computed through the use of a symplectic integrator. The cutoff parameter, $M$, in the integral is determined in the following manner.  Let $I_M$ be the calculated Casimir force up to $M$.  Now, since the integral is convergent, there exists a value of $M$ such that the contribution of the integral beyond $M$ is negligible.  In effect, this can be determined as the first value of $M$ for which $|I_{M+1} - I_{M}|<\delta$ for some arbitrary $\delta$.  Here, $\delta$ is chosen to be of the same order as the error obtained in the numerical integration of (\ref{3.1}).  This efficient process is highly accurate, allows for large flexibility, and offers improved confidence in the results obtained.

\subsection{Positive and Negative Potential}
\begin{figure}[h]
\begin{center}
\includegraphics[scale=0.85]{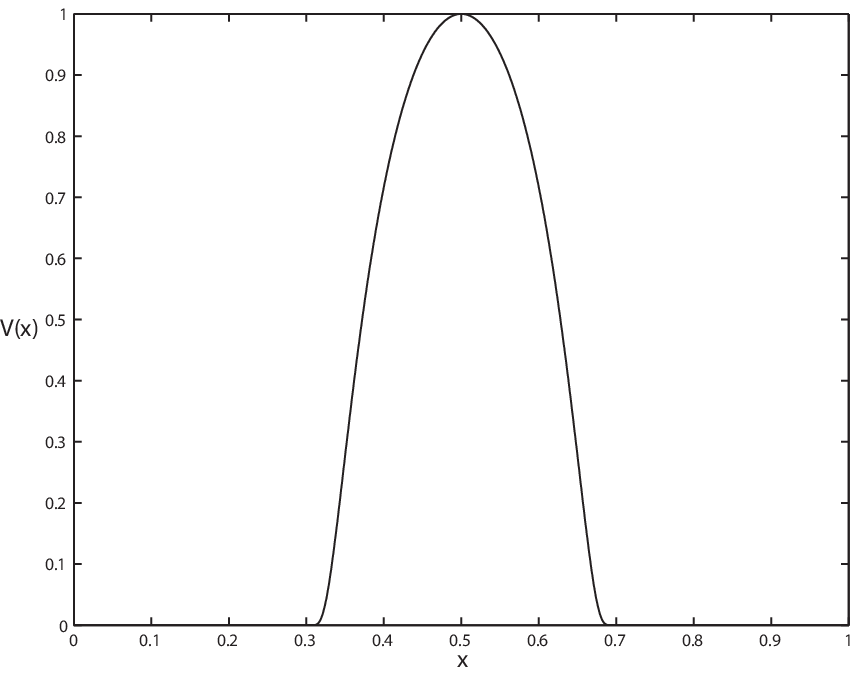}
\includegraphics[scale=0.85]{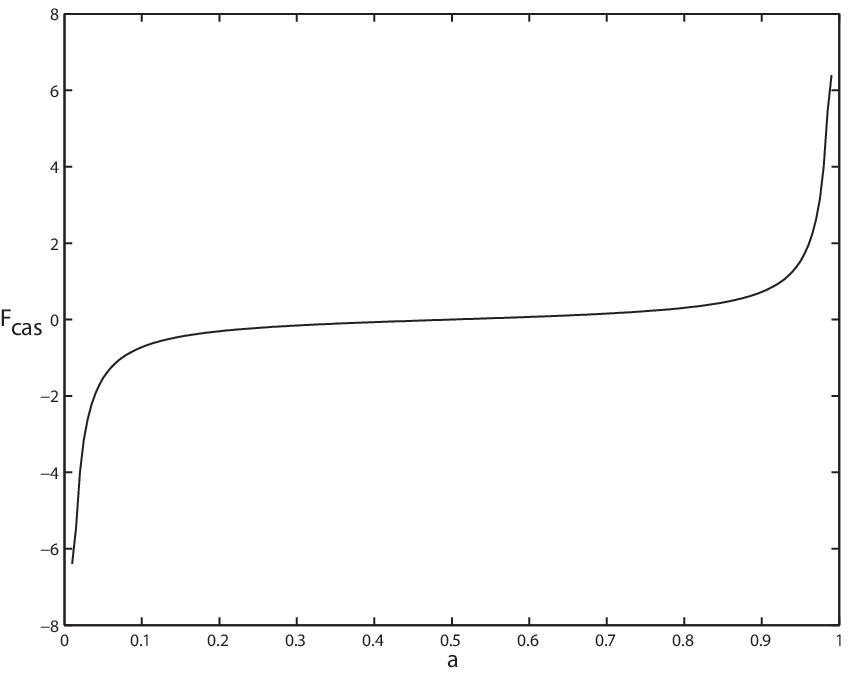}
\caption{(a) The potential is shown over the domain of its support. (b) The Casimir force, $F_{cas}$, is calculated using a standard potential of radius $\epsilon =10^{-4}$ centered at $a$ and $\eta=1$. }\label{Figure1Pot1Fcas}
\end{center}
\end{figure}

By setting $\eta=1$ in (\ref{w11}) one obtains a smooth, positive potential possessing a maximum at the point $a=1/2$. This potential is depicted in the first graph of Figure \ref{Figure1Pot1Fcas}. The Casimir force on the piston modeled by this potential is plotted in the second graph of Figure \ref{Figure1Pot1Fcas}.
The Casimir force in this case is negative when the potential is close to the left boundary at $x=0$, while it is positive when the potential is close to the right boundary at $x=1$. In addition, the force vanishes at $a=1/2$ as one would expect. This means that the positive potential considered above is always attracted to the closest boundary, making $a=1/2$ a point of unstable equilibrium.

\begin{figure}[h]
\begin{center}
\includegraphics[scale=0.85]{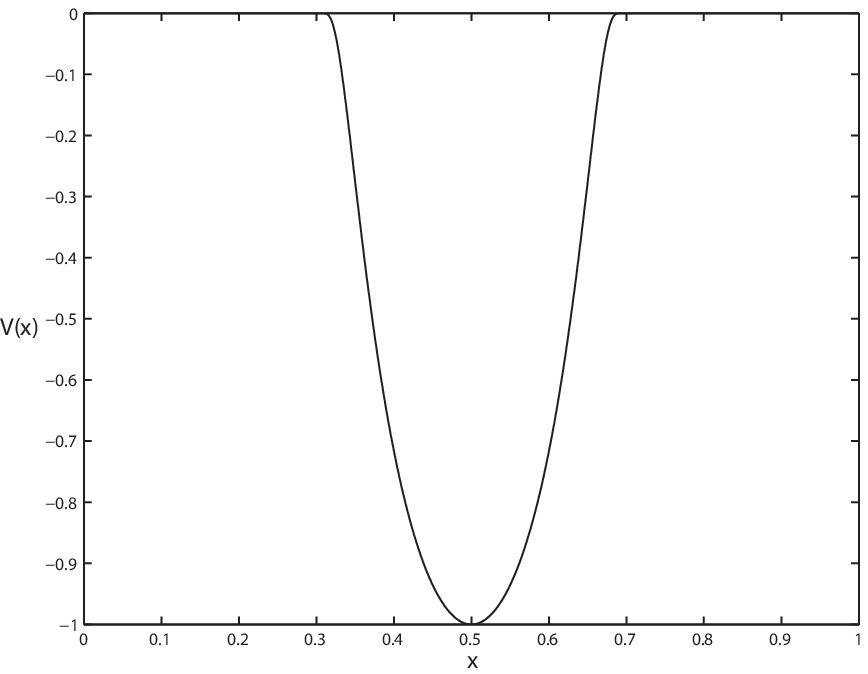}
\includegraphics[scale=0.85]{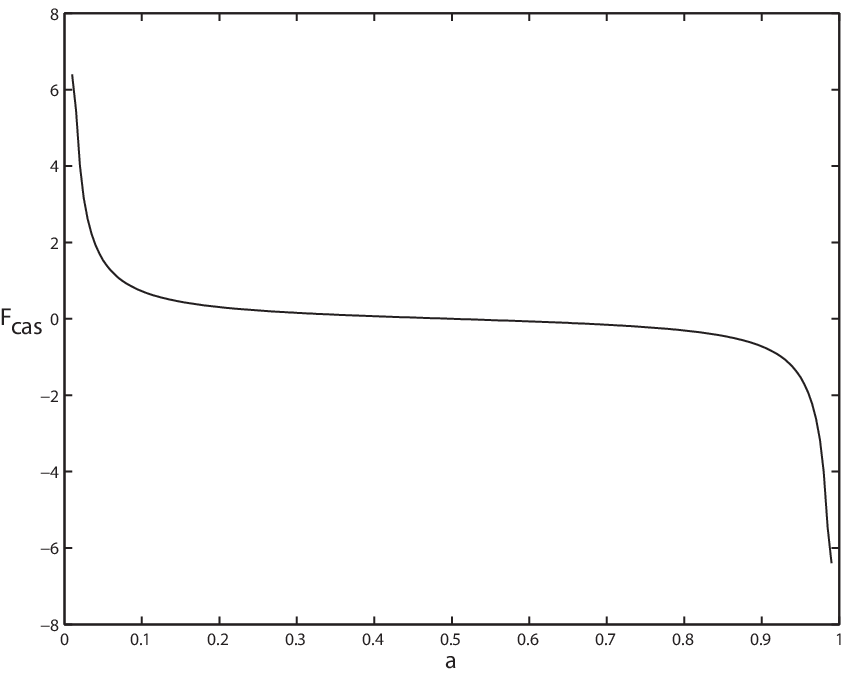}
\caption{(a) The potential is shown over the domain of its support. (b) The Casimir force, $F_{cas}$, is calculated using a standard potential of radius $\epsilon = 10^{-4}$ centered at $a$ and $\eta=-1$. }\label{Figure2Pot2Fcas}
\end{center}
\end{figure}

By setting $\eta=-1$ in (\ref{w11}) one obtains the potential illustrated in the first graph of Figure \ref{Figure2Pot2Fcas}. This potential is smooth, negative, and possesses a minimum at the point $a=1/2$. The Casimir force associated with this potential has been plotted in the second graph of Figure \ref{Figure2Pot2Fcas}.
In this situation the behavior of the Casimir force is exactly opposite to the one found for the positive potential. In particular, the negative potential is always repelled from the closest boundary and $a=1/2$ is, in this case, a point of stable equilibrium.

\subsection{Doubly-peaked Positive and Negative Potential}

\begin{figure}[h]
\begin{center}
\includegraphics[scale=0.85]{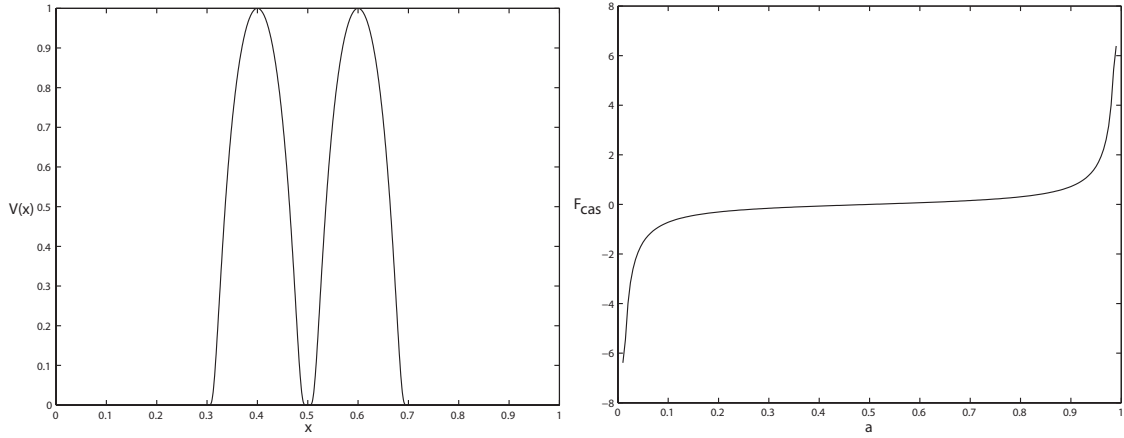}
\includegraphics[scale=0.85]{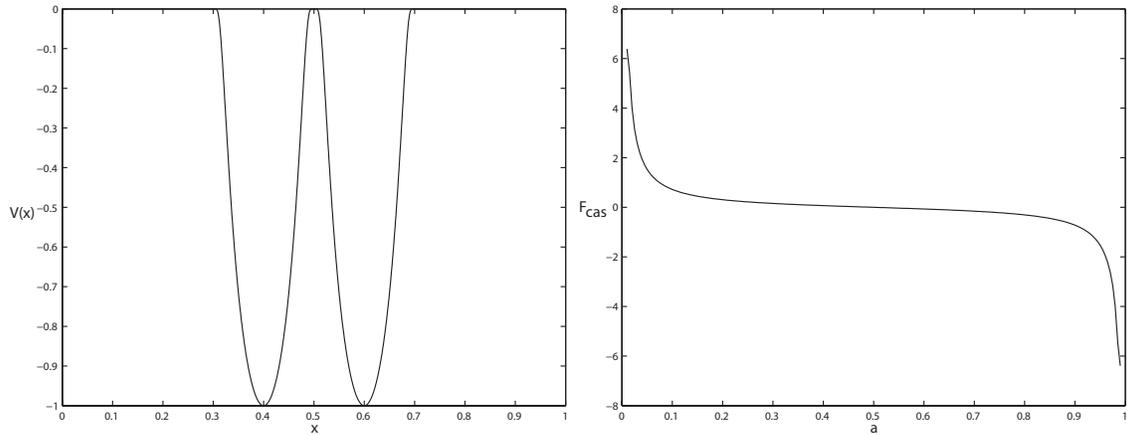}
\caption{(a) The potential is shown over the domain of its support. (b) The Casimir force, $F_{cas}$, is calculated using a double potential of radius $\epsilon = 10^{-4}$ centered at $a$. }\label{Figure3Pot5Fcas}
\end{center}
\end{figure}

The doubly-peaked positive potential is constructed by setting $\eta=1/2$ and by adding two potentials of the form (\ref{w11}) one with center at $1/2+\epsilon/2$ and the other centered at $1/2-\epsilon/2$.
The resulting potential, in the first plot of Figure \ref{Figure3Pot5Fcas}, is smooth, positive, possessing two maxima and a minimum at $a=1/2$.
The Casimir force for this potential is reported in the second graph of Figure \ref{Figure3Pot5Fcas}. The behavior of the force as
function of the position of the potential is qualitatively similar to the case of the positive potential. In particular, the doubly-peaked positive potential is
always attracted to the closest boundary.

\begin{figure}[h]
\begin{center}
\includegraphics[scale=0.85]{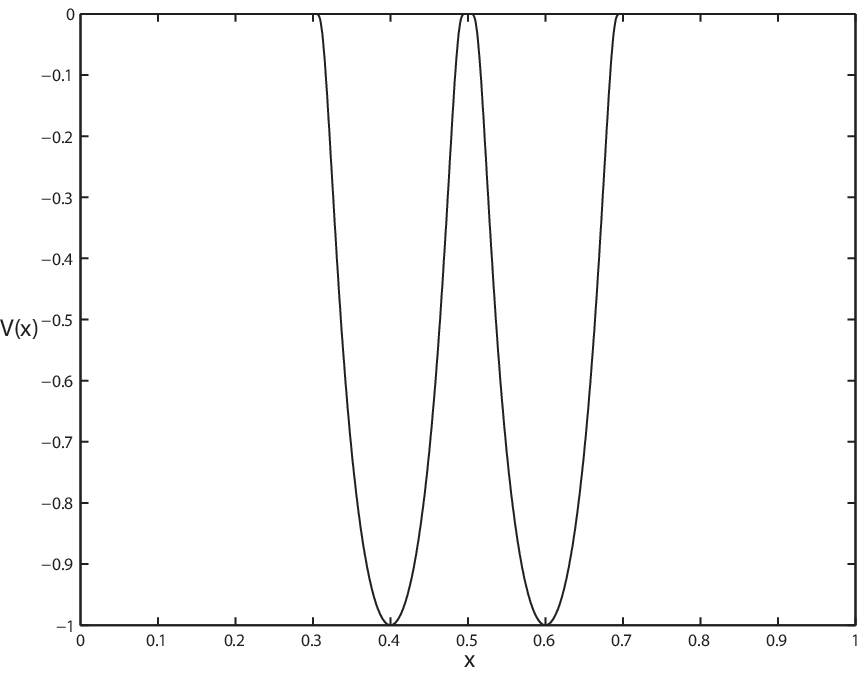}
\includegraphics[scale=0.85]{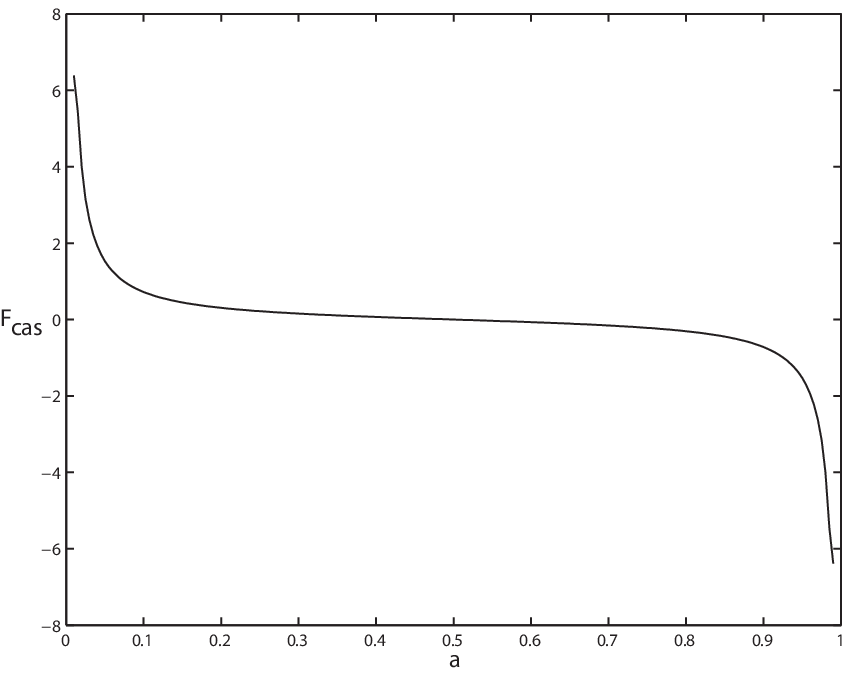}
\caption{(a) The potential is shown over the domain of its support. (b) The Casimir force, $F_{cas}$, is calculated using a double potential of radius $\epsilon = 10^{-4}$ centered at $a$. }\label{Figure4Pot6Fcas}
\end{center}
\end{figure}

The doubly-peaked negative potential is constructed in the same way as the doubly-peaked positive potential by setting $\eta=-1/2$. The resulting potential, characterized by two minima and one maximum at $a=1/2$, is depicted in the first graph of Figure \ref{Figure4Pot6Fcas} and the associated Casimir force is plotted in the second graph.
Hence, the Casimir force in this case is analogous to the one found for the negative potential. This means that the doubly-peaked negative potential is always repelled from the closest boundary.

It is not surprising that the behavior of the Casimir force on the doubly-peaked potentials matches the one found for the single potentials considered in the previous subsection. This should be expected since, effectively, the potentials have been evenly split while maintaining the total area under the curve equal to the single potentials.

\subsection{Mixed Potential}

\begin{figure}[h]
\begin{center}
\includegraphics[scale=0.85]{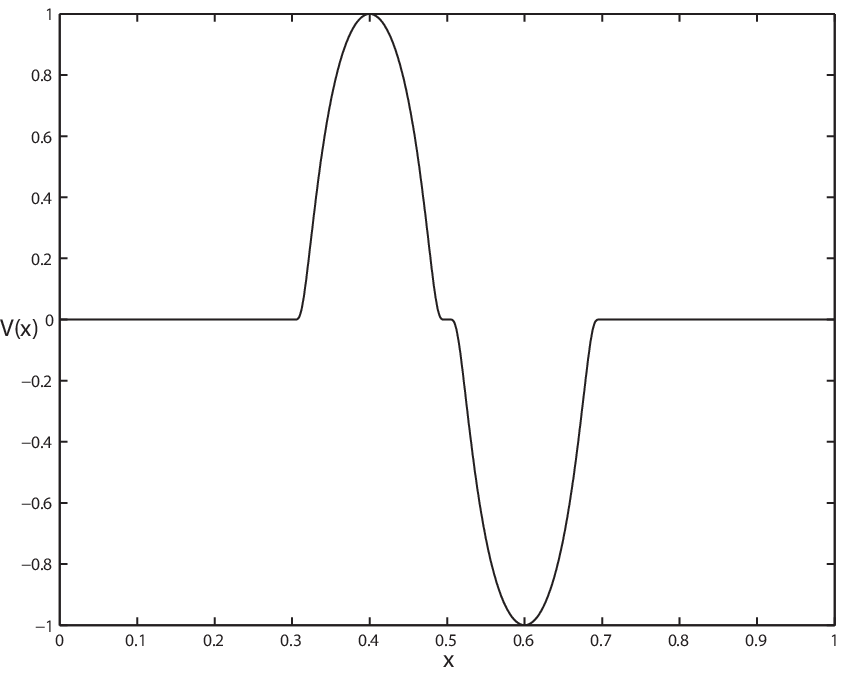}
\includegraphics[scale=0.85]{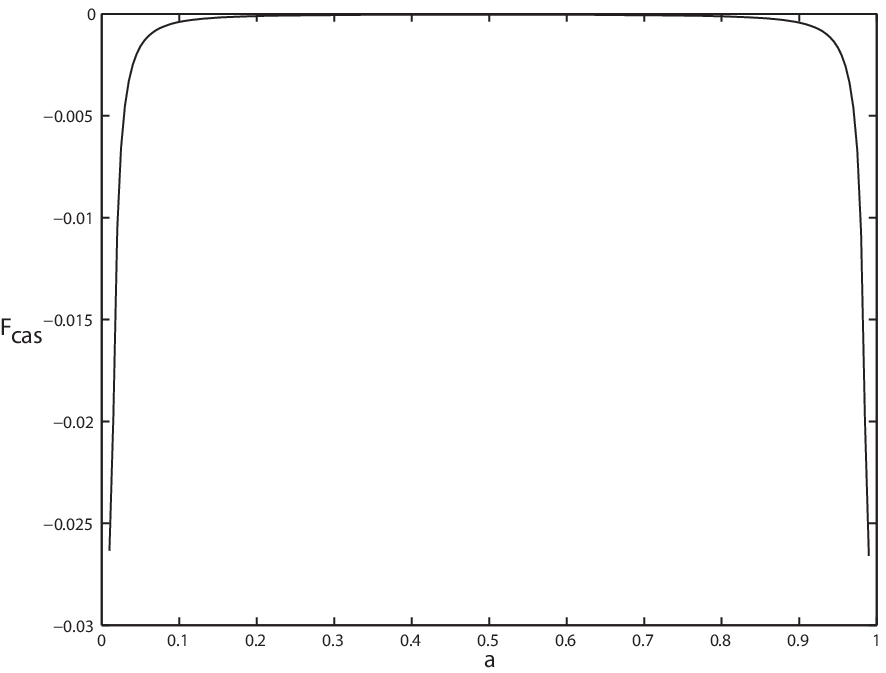}
\caption{(a) The potential is shown over the domain of its support. (b) The Casimir force, $F_{cas}$, is calculated using a double potential of radius $\epsilon = 10^{-4}$ centered at $a$. }\label{Figure5Pot3Fcas}
\end{center}
\end{figure}

The mixed potential, illustrated in the first plot of Figure \ref{Figure5Pot3Fcas}, is obtained by adding two potentials of the form (\ref{w11}): One with $\eta=1/2$ and center at $1/2-\epsilon/2$ and the other with $\eta=-1/2$, centered at $1/2+\epsilon/2$. The resulting Casimir force acting on this potential is provided by the second plot in Figure \ref{Figure5Pot3Fcas}. It is observed that the force on the potential
is always negative, in contrast to the other cases considered.  In other words, the mixed potential is repelled from the right boundary at $x=1$ but attracted to the left boundary at $x=0$.
It is interesting to notice that the opposite signs of the potential have the net effect of eliminating the force throughout the majority of the interval except for the regions that are closer to the boundary.
In the proximity of the right boundary the negative part of the potential becomes dominant and, therefore, the resulting Casimir force in that region resembles the one for negative potentials.
Near the left boundary, instead, the positive part of the potential becomes dominant resulting in a force that behaves similarly to the one found in the case of positive potentials.

\section{Higher dimensional pistons}\label{hdp}
In this section higher dimensional pistons modeled by potentials and constructed as product manifolds are studied. Let $M$ be a $D=(d+1)$-dimensional manifold such that $M=[0,L]\times {\mathcal N}$, where ${\mathcal N}$ is a smooth $d$-dimensional Riemannian manifold representing the additional Kaluza-Klein dimensions. From the manifold $M$ a piston configuration is obtained by modeling the piston itself with a smooth potential $V(x)$ having support in the interior of the interval $[0,L]$. In this setting, the manifold  $\mathscr{N}$ represents the cross-section of the piston.

The dynamics of massless scalar fields propagating on $M$ under the influence of the potential $V(x)$ is described by the operator
\begin{eqnarray}
\mathcal{L} = - \frac{\partial^2}{\partial x^2} - \Delta_{{\cal N}} + V(x)\;, \label{2.17}
\end{eqnarray}
where $\Delta_{\cal N}$ denotes the Laplacian operator on the manifold ${\cal N}$. The eigenvalue equation
\begin{equation}\label{w12}
\mathscr{L}\phi(x,y)=\lambda^{2}\phi(x,y)\;,
\end{equation}
is separable and its solutions can, hence, be written as a product
\beq\label{w13}
\phi (x,y) = X(x) \varphi (y)\;,
\eeq
where $y$ denote the coordinates on the manifold ${\cal N}$ and $\varphi (y)$ represent the eigenfunctions
of $\Delta_{\cal N}$ satisfying
\begin{eqnarray}
- \Delta _{{\cal N}} \varphi _\ell (y) = \eta_\ell ^2 \varphi _\ell (y)\;. \label{2.18}
\end{eqnarray}
By substituting the expression (\ref{w13}) in the equation (\ref{w12}), and by setting $\lambda^{2}=\nu^{2}+\eta^{2}_{\ell}$ one can show that the functions $X(x)$
are solutions to the equation
\begin{eqnarray}
\left( - \frac{d^2} {d x^2} + V(x) \right) X_\nu (x) = \nu^2 X_\nu (x)\;. \label{2.19}
\end{eqnarray}
The eigenvalues $\nu$ appearing in (\ref{2.19}) are uniquely determined once the boundary conditions for $X(x)$ have been specified. As previously done in Section \ref{onedim}, Dirichlet boundary conditions are imposed, namely,
\beq
 X_\nu (0) = X_\nu (L) =0\;.
\eeq

The spectral zeta function for the system under consideration is
\begin{eqnarray}
\zeta (s) =\sum_{\lambda} \lambda^{-2s}= \sum_{\ell, \nu} (\nu^2 + \eta_\ell^2)^{-s}\;, \label{5emi}
\end{eqnarray}
which converges for $\Re (s) > (d+1)/2$.

It is clear, at this point, that since the eigenvalue equation (\ref{w12}) is separable, leading to (\ref{2.18}) and to (\ref{2.19}), the analysis of the Casimir energy for higher dimensional pistons and the associated Casimir force can be performed by using the methods described for the one-dimensional case.

Following the ideas developed in Section \ref{onedim}, the spectral zeta function (\ref{5emi}) is represented in terms of a contour integral as follows
\beq
\zeta (s) = \frac 1 {2\pi  i} \sum_\ell \int\limits_\gamma d\mu (\mu^2 + \eta_\ell ^2)^{-s} \frac d {d\mu} \ln u_\mu (L)\;,
\eeq
where $u_{\mu}(x)$ are the solutions to the initial value problem (cf. Section \ref{onedim})
\begin{equation}
  \left(-\frac{d^2}{dx^{2}}+V(x)\right)u_{\mu}(x)=\mu^{2}u_{\mu}(x)\;,\qquad u_{\mu}(0)=0\;,\quad u^{\prime}_{\mu}(0)=1\;.
\end{equation}

Due to the presence of the manifold ${\cal N}$ the analytically continued expression for the spectral zeta function $\zeta(s)$ will be written in terms of the spectral zeta function
associated with $\Delta_{\cal N}$, namely
\beq
\zeta_{{\cal N}} (s) = \sum_\ell \eta_\ell ^{-2s}\;,
\eeq
which is well defined for $\Re (s) > d/2$ and can be extended to a meromorphic function in the entire complex plane possessing only simple poles.
The standard technique of adding and subtracting asymptotic terms is performed, in particular, \cite{fucci12}
\beq & &\hspace{-.5cm}\zeta ^{(f)}(s) = \frac{\sin \pi s} \pi \sum_\ell \int\limits_{\eta _\ell}^\infty dk \,\, (k^2 - \eta_\ell^2)^{-s} \frac d {dk} \left\{ \ln u_{ik} (L) - kL + \ln (2k) -\sum_{j=0}^N \frac{d_j}{ k^{j}} \right\} , \label{2.20}\\
 & &\hspace{-.5cm}\zeta ^{(as)} (s) = \frac 1 {2 \Gamma (s)} \left\{  \frac{ L \Gamma \left( s- \frac 1 2 \right)}{\sqrt \pi} \zeta _{{\cal N}} \left( s- \frac 1 2 \right)
- \Gamma (s) \zeta _{{\cal N}} (s) \right.\nn\\
& &\hspace{3.0cm}\left.- \sum_{j=1}^N j d_j \frac{ \Gamma \left( s+ \frac j 2 \right)}{\Gamma \left( 1 + \frac j 2 \right)} \zeta_{{\cal N}} \left( s + \frac j 2\right) \right\}.\label{2.21}\eeq
Here, $\zeta^{(f)}(s)$ can be proved to be well defined for $(D-N-2)/2<\Re (s)<1$. By choosing $N=D$, $\zeta^{(f)}(s)$ becomes an analytic function in the neighborhood of $s=-1/2$ and can, therefore, be used for the computation of the Casimir energy with no further manipulations. Using the well-known meromorphic structure of the spectral zeta function associated with the Laplacian on $\mathscr{N}$ \cite{kirs02b}, the relevant expression for $\zeta(s)$ at $s=-1/2$ reads
\beq\label{w22}
\zeta ^{(f)} (-1/2) &=& - \frac 1 \pi \sum_\ell \int\limits_{\eta_\ell } ^\infty dk \,\, (k^2 - \eta_\ell^2)^{1/2} \frac d {dk} \left\{ \ln u_{ik} (L) - kL + \ln (2k) - \sum_{j=1}^D \frac{d_j}{ k^{j}} \right\}, \;\;\;\;
\eeq
and
\begin{eqnarray}\label{w23}
\zeta ^{(as)} (-1/2+\epsilon) &=& \frac 1 \epsilon \Bigg\{ \frac L {4\pi} \zeta_{{\cal N}} (-1) - \frac 1 2 \mbox{Res } \zeta_{{\cal N}} (-1/2) + \frac{d_1 \zeta_{{\cal N}} (0)} {2\pi} \nn\\
&+& \sum_{j=2}^D \frac{d_j}{2\sqrt \pi} \frac{\Gamma \left( \frac{j-1} 2 \right)} {\Gamma \left( \frac j 2 \right)} \mbox{Res } \zeta_{{\cal N}} \left( \frac{j-1} 2 \right) \Bigg\}\nn\\
&+& \frac L {4\pi} \left[ \zeta_{{\cal N}} ' (-1) + \zeta_{{\cal N}} (-1) \left( \ln 4-1\right) \right]- \frac 1 2 \mbox{FP } \zeta_{{\cal N}} (-1/2) \nn\\
&+&\frac{d_1}{2\pi} \left[ \zeta_{{\cal N}}  ' (0) + \zeta_{{\cal N}} (0) \left(\ln 4-2\right)\right]+\sum_{j=2}^D \frac{ d_j} {2\sqrt \pi} \frac{\Gamma \left( \frac{j-1} 2 \right)} {\Gamma \left( \frac j 2 \right) } \Bigg[ \mbox{FP} \zeta_{{\cal N}} \left( \frac{j-1} 2 \right) \nn\\
&+& \mbox{Res}\zeta_{{\cal N}} \left( \frac{j-1} 2 \right) \left(\Psi\left(\frac{j-1}{2}\right)-\gamma-\ln 4+2\right)\Bigg]+O(\epsilon)\;,
\end{eqnarray}
where $\Psi(x)$ represents the logarithmic derivative of the Euler gamma function.
According to (\ref{2.4}), the Casimir energy is obtained by adding the expressions (\ref{w22}) and (\ref{w23}) and by multiplying by $1/2$. It is worth noting that
the above results are quite general and valid for an arbitrary smooth Riemannian manifold $\mathscr{N}$ and for any smooth potential with compact support in $[0,L]$. More explicit results can only be found once the manifold $\mathscr{N}$ and the potential $V(x)$ are completely specified. Despite the lack of an explicit expression for the Casimir energy, one can still make some general remarks. From the results (\ref{w22}) and (\ref{w23}) it is clear that the Casimir energy is, generally, divergent. However, since none of the terms in (\ref{w23}) depend on the variable $a$, the resulting Casimir force acting on the piston acquires contributions only from the finite part in (\ref{w22}) and is, hence, finite.

  This section will be concluded by discussing two specific examples. Consider the particular cases for which ${\cal N} =\reals$ and ${\cal N}=\reals^2,$ respectively. In such cases the continuous spectrum resulting from the unrestricted dimensions can be integrated to obtain the following spectral zeta function densities
\beq
\zeta _{I\times \reals} (s) = \frac{\Gamma \left( s-\frac 1 2 \right)}{\sqrt {4\pi} \Gamma (s) } \zeta_I \left( s- \frac 1 2 \right)\;,
\eeq
respectively
\beq
\zeta_{I\times \reals^2} (s) = \frac 1 {4\pi (s-1)} \zeta _I (s-1)\;.
\eeq
Using the known meromorphic structure of $\zeta _I (s)$ it is then easily verified that
\beq
F_{Cas} ^{I\times \reals} = - \frac 1 {8\pi} \frac \partial {\partial a} \zeta_I ' (-1)
= \frac 1 {8\pi } \int\limits_m^\infty dk (k^2-m^2) \frac \partial {\partial a} \frac \partial {\partial k} \ln u_{ik} (L)\;,
\eeq
and
\beq
F_{Cas} ^{I\times \reals^2}= \frac 1 {12\pi} \frac \partial {\partial a} \mbox{FP }\zeta_I  \left(-\frac 3 2\right)= \frac 1 {12\pi } \int\limits_m^\infty dk (k^2-m^2)^{3/2} \frac \partial {\partial a} \frac \partial {\partial k} \ln u_{ik} (L)\;,
\eeq
where, again, $\mbox{FP}$ denotes the finite part of the Laurent series expansion. Numerical results for the above expressions can be obtained by following the same procedure described in the previous section.
When ${\cal N}$ is either $\mathbb{R}$ or $\mathbb{R}^{2}$ the Casimir force on pistons modeled by the type of potentials
described in Section \ref{Sec:Examples} is qualitatively similar to the force found in one dimension for the same type of potentials.
Since in this case the plots of the force as a function of the position $a$ resemble the ones in Figures \ref{Figure1Pot1Fcas}-\ref{Figure5Pot3Fcas}(b), for brevity, they are not included here.

\section{Conclusions}

In this paper a method is developed to study the Casimir energy for massive scalar fields confined in finite volumes under the influence of smooth background potentials.
The pivotal point of our method relies on rewriting the eigenvalues of the relevant {\it boundary value problem} as solutions to a transcendental equation related to an equivalent {\it initial value problem}; see Equations (\ref{2.5}) and (\ref{2.6}). The starting point of our approach is the representation of the spectral zeta function associated with our models in terms of a complex integral. The analytic continuation of $\zeta(s)$ to a neighborhood of $s=-1/2$ was then achieved by adding and subtracting a suitable number of asymptotic terms from the integral representation. Although the form of the potential has been left unspecified, standard WKB techniques have allowed us to effectively compute the asymptotic expansions needed for the analytic continuation. The obtained analytic continuation of the spectral zeta function is, then, used in order to compute the Casimir force acting on a piston modeled by a smooth potential with compact support. In this case, it is found that while the Casimir energy is, in general, divergent the Casimir force on the piston is well defined. This can be immediately understood by noticing that the divergent terms in the energy
are independent of the position of the piston.

In the framework of pistons modeled by potentials it is found that the Casimir force can only be computed numerically once a potential has been specified.
For this reason, several types of Gaussian potentials have been considered and plots of the Casimir force on the piston as a function of the position $a$ have been provided.
We would like to stress, at this point, that in our examples Gaussian potentials have been chosen for illustrative purposes only.
In fact, the results obtained in this work are completely general and are valid for {\it any} smooth potential with compact support.
The numerical results obtained for the Casimir force in the various examples are consistent with our physical intuition, a fact that improves confidence in the presented analysis.

This work can naturally be continued by considering the effect that different types of boundary conditions, such as Neuman, Robin, or mixed, have on the Casimir force. This analysis would follow the lines presented here without any major technical complications. In fact, the boundary conditions determine uniquely the constants $A^{+}$ and $A^{-}$ in the asymptotic expansion of the functions in (\ref{w3}). Different types of boundary conditions will lead to different expressions for $A^{+}$ and $A^{-}$ but will keep the form of the asymptotic expansion (\ref{w6}) unchanged. The analytic continuation of the spectral zeta function would, then, proceed in the same way as presented in this work.

In addition, of particular interest is to study higher dimensional pistons modeled by potentials when the additional Kaluza-Klein manifold $\mathscr{N}$ allows for the explicit knowledge of the eigenvalue associated with $\Delta_{\mathscr{N}}$. Along these lines the authors are currently in the process of analyzing spherically symmetric and cylindrically symmetric configurations where angular momentum sums introduce additional technical and numerical complications.


\begin{thebibliography}{10}

\bibitem{acto95-52-3581}
A.A. Actor and I.~Bender.
\newblock Casimir effect for soft boundaries.
\newblock {\em Phys. Rev.}, D52:3581--3590, 1995.

\bibitem{adki83-228-552}
G.S. Adkins, C.R. Nappi, and E.~Witten.
\newblock Static properties of nucleons in the {S}kyrme model.
\newblock {\em Nucl. Phys.}, B228:552--566, 1983.

\bibitem{ambj85-256-434}
J.~Ambjorn and V.A. Rubakov.
\newblock Classical versus semiclassical electroweak decay of a techniskyrmion.
\newblock {\em Nucl. Phys.}, B256:434--448, 1985.

\bibitem{bend10b}
C.~Bender and S.~Orszag
\newblock {\em Advanced mathematical methods for scientists and engineers I:
  Asymptotic methods and perturbation theory}.
\newblock Springer, New York, 1999.

\bibitem{bord96-53-5753}
M.~Bordag and K.~Kirsten.
\newblock Vacuum energy in a spherically symmetric background field.
\newblock {\em Phys. Rev.}, D53:5753--5760, 1996.

\bibitem{bord09b}
M.~Bordag, G.L. Klimchitskaya, U.~Mohideen, and V.M. Mostepanenko.
\newblock {\em Advances in the Casimir effect}.
\newblock Oxford Science Publications, 2009.

\bibitem{bord01-353-1}
M.~Bordag, U.~Mohideen, and V.M. Mostepanenko.
\newblock New developments in the {C}asimir effect.
\newblock {\em Phys. Rept.}, 353:1--205, 2001.

\bibitem{casi48-51-793}
H.B.G. Casimir.
\newblock On the attraction between two perfectly conducting plates.
\newblock {\em Kon. Ned. Akad. Wetensch. Proc.}, 51:793--795, 1948.

\bibitem{cava04-69-065015}
R.M. Cavalcanti.
\newblock {Casimir force on a piston}.
\newblock {\em Phys. Rev.}, D69:065015, 2004.


\bibitem{conw78b}
J.~Conway.
\newblock {\em Functions of one Complex Variable}.
\newblock Springer-Verlag, New York, 1978.

\bibitem{dunn06-39-11915}
G.V. Dunne and K.~Kirsten.
\newblock {Functional determinants for radial operators}.
\newblock {\em J. Phys.}, A39:11915--11928, 2006.

\bibitem{dunn09-42-075402}
G.V. Dunne and K.~Kirsten.
\newblock {Simplified Vacuum Energy Expressions for Radial Backgrounds and
  Domain Walls}.
\newblock {\em J. Phys. A: Math. Theor.}, 42:075402, 2009.

\bibitem{eila86-56-1331}
G.~Eilam, D.~Klabucar, and A.~Stern.
\newblock Skyrmion solutions to the {W}einberg-{S}alam model.
\newblock {\em Phys. Rev. Lett.}, 56:1331--1334, 1986.

\bibitem{eliz95b}
E.~Elizalde.
\newblock {\em Ten Physical Applications of Spectral Zeta Functions}.
\newblock Lecture Notes in Physics m35, Springer-Verlag, Berlin, 1995.

\bibitem{eliz09-79-065023}
E.~Elizalde, S.D. Odintsov, and A.A. Saharian.
\newblock {Repulsive Casimir effect from extra dimensions and Robin boundary
  conditions: from branes to pistons}.
\newblock {\em Phys. Rev.}, D79:065023, 2009.

\bibitem{eliz97-30-5393}
E.~Elizalde and A.~Romeo.
\newblock One-dimensional Casimir effect perturbed by an external field.
\newblock {\em J. Phys. A: Math. Gen.}, 30:5393--5403, 1997.

\bibitem{fucci12}
G.~Fucci, K.~Kirsten, and P.~Morales.
\newblock Pistons modeled by potentials.
\newblock in {\it Cosmology, Quantum Vacuum, and Zeta
Functions}, eds. S. Odintsov, D. S\'{a}ez-G\'{o}mez, and S. Xamb\'{o}, (Springer-Verlag, Berlin, 2011) p. 313-322.

\bibitem{fucci12a}
G.~Fucci and K.~Kirsten.
\newblock The Casimir effect for generalized piston geometries.
\newblock {\em Int. J. Mod. Phys. A}, 27:1260008 (2012)


\bibitem{frie77-15-1694}
R.~Friedberg and T.D. Lee.
\newblock Fermion field nontopological solitons. {I}.
\newblock {\em Phys. Rev.}, D15:1694--1711, 1977.

\bibitem{frie77-16-1096}
R.~Friedberg and T.D. Lee.
\newblock Fermion field nontopological solitons. {I}{I}. {M}odels for hadrons.
\newblock {\em Phys. Rev.}, D16:1096--1118, 1977.

\bibitem{gips84-231-365}
J.M. Gipson.
\newblock Quasi-solitons in the strongly coupled {H}iggs sector of the standard
  model.
\newblock {\em Nucl. Phys.}, B231:365--385, 1984.

\bibitem{gips81-183-524}
J.M. Gipson and C.-H. Tze.
\newblock Possible heavy solitons in the strongly coupled {H}iggs sector.
\newblock {\em Nucl. Phys.}, B183:524--546, 1981.

\bibitem{hert05-95-250402}
M.P. Hertzberg, R.L. Jaffe, M.~Kardar, and A.~Scardicchio.
\newblock {Attractive Casimir Forces in a Closed Geometry}.
\newblock {\em Phys. Rev. Lett.}, 95:250402, 2005.

\bibitem{kirs02b}
K.~Kirsten.
\newblock {\em Spectral Functions in Mathematics and Physics}.
\newblock Chapman\&Hall/CRC, Boca Raton, FL, 2002.

\bibitem{kirs09-79-065019}
K.~Kirsten and S.A. Fulling.
\newblock Kaluza-{K}lein models as pistons.
\newblock {\em Phys. Rev.}, D79:065019, 2009.

\bibitem{kirs03-308-502}
K.~Kirsten and A.J. McKane.
\newblock Functional determinants by contour integration methods.
\newblock {\em Ann. Phys.}, 308:502--527, 2003.

\bibitem{kirs04-37-4649}
K.~Kirsten and A.J. McKane.
\newblock Functional determinants for general {S}turm-{L}iouville problems.
\newblock {\em J. Phys. A: Math. Gen.}, 37:4649--4670, 2004.

\bibitem{klin84-30-2212}
F.R. Klinkhamer and N.S. Manton.
\newblock A saddle point solution in the {W}einberg-{S}alam theory.
\newblock {\em Phys. Rev.}, D30:2212--2220, 1984.

\bibitem{mara07-75-085019}
V.~Marachevsky.
\newblock {Casimir interaction of two plates inside a cylinder}.
\newblock {\em Phys. Rev.}, D75:085019, 2007.

\bibitem{mill06b}
P.D. Miller.
\newblock {\em Applied asymptotic analysis}.
\newblock American Mathematical Society, Providence, Rhode Island, 2006.

\bibitem{milo94b}
P.W. Milonni.
\newblock {\em The Quantum Vacuum: An Introduction to Quantum Electrodynamics}.
\newblock Academic Press, New York, 1994.

\bibitem{milt01b}
K.A. Milton.
\newblock {\em The {C}asimir Effect: Physical Manifestations of Zero-Point
  Energy}.
\newblock River Edge, USA: World Scientific, 2001.

\bibitem{most97b}
V.M. Mostepanenko and N.N. Trunov.
\newblock {\em The {C}asimir Effect and Its Applications}.
\newblock Clarendon, Oxford, 1997.

\bibitem{poly74-20-194}
A.M. Polyakov.
\newblock Particle spectrum in quantum field theory.
\newblock {\em JETP Lett.}, 20:194--195, 1974.

\bibitem{skyr61-260-127}
T.H.R. Skyrme.
\newblock A nonlinear field theory.
\newblock {\em Proc. Roy. Soc. Lond.}, A260:127--138, 1961.

\bibitem{skyr62-31-556}
T.H.R. Skyrme.
\newblock A unified field theory of mesons and baryons.
\newblock {\em Nucl. Phys.}, B31:556--569, 1962.

\bibitem{hoof74-79-276}
G.'t~Hooft.
\newblock Magnetic monopoles in unified gauge theories.
\newblock {\em Nucl. Phys.}, B79:276--284, 1974.


\end{thebibliography}


\end{document}